# Electronic inhomogeneity at magnetic domain walls in strongly-correlated systems

M.S. Rzchowski and Robert Joynt

*Physics Department, University of Wisconsin-Madison*
*Madison, WI 53706*

We show that nano-scale variations of the order parameter in strongly-correlated systems can induce local spatial regions such as domain walls that exhibit electronic properties representative of a different, but nearby, part of the phase diagram. This is done by means of a Landau-Ginzburg analysis of a metallic ferromagnetic system near an antiferromagnetic phase boundary. The strong spin gradients at a wall between domains of different spin orientation drive the formation of a new type of domain wall, where the central core is an insulating antiferromagnet, and connects two metallic ferromagnetic domains. We calculate the charge transport properties of this wall, and find that its resistance is large enough to account for recent experimental results in colossal magnetoresistance materials. The technological implications of this finding for switchable magnetic media are discussed.

There is growing evidence that electronic spatial inhomogeneity on the nanoscale is an intrinsic property of strongly-correlated oxide materials. In one sense, this is not surprising, since the strong competing interactions produce many phases whose free energies are nearly degenerate, resulting in rich phase diagrams. Nevertheless, in particular cases, the origin of the phenomenon is often unclear, and its appearance is usually unpredictable and uncontrollable. The influence of this inhomogeneity on the macroscopic properties can be profound.

A very important example is the family of high-$T_c$ superconductors. The proximity of antiferromagnetic (AF) and conducting phases has led many authors to propose spatially inhomogeneous states, usually with a striped configuration [1]. For some materials and doping concentrations, there is unambiguous evidence for these inhomogeneities [2]. One may extend these ideas and propose that the cores of vortices, which are *induced* inhomogeneities, may be antiferromagnetic, i.e., may show signs of the proximity of an insulating phase in the doping-temperature phase diagram [3].

Another important class of oxides, the colossal magnetoresistive (CMR) materials, have very solid evidence for electronic inhomogeneity. Transmission electron microscopy has indicated [4] spatial charge separation, and scanning tunneling microscopy measurements [5] have provided strong evidence for a temperature range of local electronic phase separation even in nominally homogeneous ferromagnetic (FM) CMR materials. Measurements of electronic transport through magnetic domain walls in single-crystalline CMR materials [6] indicate a domain-wall resistance orders of magnitude greater than that attributable to scattering from the spin gradient in a Bloch wall, which is generally a very small effect [7,8]. We propose in this Letter that this phenomenon is the result of an AF core in the domain wall. Since the AF phase is insulating, the effect on transport is dramatic. This has important technological implications. Domain-wall populations in films can be controlled with relatively small magnetic fields: they can be induced, as well as natural, inhomogeneities. Thus very large controllable magnetoresistances can be envisioned. This is a central requirement for magnetic memory elements.

The bulk phase diagram in the temperature-doping plane of the CMR materials indicates a transition, usually first-order, from a metallic FM phase to an insulating AF phase. In the FM phase the order parameter (OP) has degenerate configurations, and domains of different ground states occur. In a domain wall, where the OP interpolates between the two configurations in a narrow region, there are strong gradients, and the bulk phase diagram no longer applies.

If the system is close to a bulk phase boundary, it can be energetically favorable for a region of the wall to collapse into a different electronic phase rather than supporting a strongly rotating OP. Our calculations show that this can have drastic effects. It is possible to change the character of the domain wall from metallic to insulating, or even to dramatically change the spatial extent of the wall. As the system approaches the phase boundary, the width of the wall can diverge, or the walls could proliferate. This growth in spatial extent could be the origin of electronic phase separation observed in some of these materials. This would clearly affect macroscopic properties when the domain walls make up a significant fraction of the sample. Thus our theory of domain walls in films tends to confirm theories of bulk CMR which depend on the existence of mesoscopic patterns in which metallic and insulating regions interpenetrate [9].

Here we investigate the case of a single domain walls between FM domains in a material close to an AF phase boundary. However, the theoretical part of the paper is more general, and would apply to domain walls in other systems with closely competing OP's.

II. Electronic phase of the domain wall

We model this system with a one-dimensional Landau-Ginzburg free energy

$$F = -a_m \vec{M}^2 + b_m \vec{M}^4 - cM_z^2 + g_m(\vec{M}')^2 - a_a A^2 + b_a A^4 + g_a(A')^2 + gA^2\vec{M}^2 \quad (1)$$

Here $\vec{M}$ is a vector OP for ferromagnetism, and $A$ is a scalar OP representing the competing antiferromagnetism. Primes denote spatial derivatives. $A$ is a scalar only for the sake of simplicity. The results for vector $A$ would be very similar. We consider temperatures well below the FM ordering temperature, where all coefficients in Eq. (1) are positive. Furthermore, we are in a doping range where the bulk state is FM, but close to the 1$^{st}$-order boundary to the AF state: $0 < f \equiv F_A^{(u)}/F_M^{(u)} < 1$, where $F_A^{(u)}$ [$F_M^{(u)}$] are the bulk, uniform, free energy densities of the AF [F] states in the absence of the other. The anisotropy term $-cM_z^2$ favors alignment of the ferromagnetic moment in the $\pm\hat{z}$ direction. This is the appropriate anisotropy for the PrSrMnO$_3$ films that we will be concerned with below. The coexistence term $gA^2\vec{M}^2$ increases the free energy when the two order parameters are nonzero at the same spatial location. We will consider mainly the case where $g_A \ll g_M$. This is physically reasonable in the CMR systems since FM arises from the rather nonlocal double exchange mechanism.

We impose a magnetic domain wall by enforcing the boundary conditions $\vec{M} \to \pm M^{(u)}\hat{z}$ at $x = \pm\infty$. The wall itself itself is only thermodynamically stable in the presence of long-range (dipole or other) forces [9]. Here $M^{(u)} = ((a_m + c)/2b_m)^{1/2}$ is the equilibrium FM magnetization of the uniform, single-domain system. We explore broad regions of parameter space using simulated annealing of the free energy in Eq. (1), and present analytic solutions for particular regions of phase space.

With no coupling of the two OP's ($g = 0$), Eq. (1) describes a ferromagnetic Bloch wall. Then $g_m$ defines the spatial length scale, and $b_m$ becomes a scale factor for the overall free energy. The qualitative characteristics of the wall are controlled only by $a_m$. For $a_m$ large the magnitude of the magnetization vector is stiff, and Eq. (1) can be analytically minimized to obtain the known result for a Bloch domain wall where $\vec{M}$ rotates at constant magnitude. This gives $\theta = \arctan(\exp(x/\xi))$, where $\theta$ is the angle of the local magnetization with respect to the quantization axis, and $\xi = (2g_m/c)^{1/2}$. For smaller $a_m$, the magnitude $M$ is reduced at the wall center, and also $\vec{M}$ completes its rotation over a slightly shorter distance.

Nonzero $g$ results in competition between the two order parameters. In this case we find an additional solution for large enough $g$, shown in Fig. 1 (for $g_A = 0$ and $g = 8$), where an antiferromagnetic region (dashed line) exists in the core of the domain wall. In this case the ferromagnetic magnetization $M$ (solid lines) goes continuously through zero. The general characteristics of this type of solution are robust over a wide parameter range: $\vec{M}$ does not rotate continuously but rather is directed along $\pm\hat{z}$, changing orientation discontinuously as the magnitude passes through zero. Solutions for two different parameter sets are shown: the 'narrow' wall corresponds to parameters such that $f = F_A^{(u)}/F_F^{(u)} = 0.64$, (relatively far from the bulk phase boundary), while the 'wide' wall has $f = F_A^{(u)}/F_F^{(u)} = 0.98$ (close to the boundary). The absence of an antiferromagnetic gradient term ($g_A = 0$) leads to a discontinuous slope as $A$ goes to zero. This sharp change allows a straightforward definition of the antiferromagnetic core width as the region over which $A$ is non-zero. Figure 2 shows that this width diverges logarithmically as $f \to 1$.

These solutions have $M_y = 0$ and $M_z^2 = \vec{M}^2$. In this case, the free energy can be expressed in normalized variables to clarify the physics. Equation (1) becomes

$$\tilde{F} = f^{-1/2}[-\tilde{M}^2 + \tilde{M}^4/2 + (\tilde{M}')^2/2] + f^{1/2}[-\tilde{A}^2 + \tilde{A}^4/2] + \gamma\tilde{A}^2\tilde{M}^2, \tag{2}$$

where $\tilde{M} \equiv M_z/M^{(u)}$, $\tilde{A} = A/A^{(u)}$, $f = F_A^{(u)}/F_M^{(u)}$, $\tilde{F} = F/(F_M^{(u)} F_A^{(u)})^{1/2}$, $\gamma = g/(4b_M b_A)^{1/2}$, and the normalized spatial coordinate $\tilde{x} \equiv x/\xi_o$ with $\xi_o = (2g_m/(a_m+c))^{1/2}$. The shows that the free energy, and hence the physical behavior, is determined by only two parameters: the ratio of the free energies for the isolated uniform solutions and the coexistence energy.

This form permits an analytic calculation of the domain wall structure. Making Eq. (2) stationary with respect to variations in $\tilde{A}$ and $\tilde{M}$ leads to

$$\tilde{A}^2 = 1 - (\gamma/f^{1/2})\tilde{M}^2$$
$$\tilde{M}''/2 = \tilde{M}^3 - \tilde{M}(1 - f^{1/2}\gamma\tilde{A}^2)$$
(3)

Equation (3) shows that $\tilde{A}$ decreases as $\tilde{M}^2$ increases from zero and, for $f^{1/2}/\gamma$ large enough, $\tilde{A} = 0$ beyond a critical ferromagnetic magnetization value $\tilde{M}^2 = f^{1/2}/\gamma$. This divides the solutions to Eq. (3) into one valid in the 'exterior' ($\tilde{A} = 0$) region and one valid in the "interior" ($\tilde{A} \neq 0$). In the exterior region, Eq. (3b) can be integrated (using the boundary condition $\tilde{M}' \to 0$ as $\tilde{M} \to 1$) to find $\tilde{M}^{\text{ext}} = \tanh(\tilde{x} - \tilde{x}_o)$. $\tilde{x}_o$ is determined by matching to the interior solution. The interior solution can be integrated analytically ( after substituting for $\tilde{A}$ and matching the interior and exterior solution derivatives at $\tilde{M}^2 = f^{1/2}/\gamma$ ) to obtain

$$(\tilde{M}')^2 = (1 - \tilde{M}^2)^2 - f(1 - (\gamma/f^{1/2})\tilde{M}^2)^2,$$

which can then be integrated numerically to find $\tilde{x}(\tilde{M})$ as

$$\tilde{x} = \int_0^{\tilde{M}} \frac{d\tilde{M}}{\sqrt{(1-\tilde{M}^2)^2 - f(1-(\gamma/f^{1/2})\tilde{M}^2)^2}}.$$
(4)

The width of the domain wall antiferromagnetic core is obtained by integrating up to $\tilde{M} = (f^{1/2}/\gamma)^{1/2}$, where $\tilde{A}$ goes to zero, and multiplying by two. For large $\gamma$ the wall width diverges as $f \to 1$, which in Eq. (4) arises from the divergence of the integrand at $\tilde{M} = 0$, $f \to 1$. The leading-order behavior of Eq. (4), obtained by approximating the integrand for small $\tilde{M}$, is $-(1/\alpha)\ln(1-f) + (2/\alpha)\ln\alpha$, with $\alpha = \sqrt{2(f^{1/2}\gamma - 1)}$. This reproduces the logarithmic dependence on $1-f$, and the $\sim \gamma^{-1/2}$ dependence of the slope, shown in the inset to Fig. 2.

From a simulated annealing analysis, we find that over a broad parameter range this antiferromagnetic core solution is very close to the one that minimizes the general free energy (Eq. (1)). This is illustrated by the 'domain wall phase diagram' of Fig. 3, where the minimum free energy solutions obtained by simulated annealing of Eq. (1) are characterized for different values of the parameters $f$ and $\gamma$. The region labeled 'AF-core' is a region of parameter space where the antiferromagnetic OP $A$ is nonzero in the domain wall. Solutions in the 'F-core' region have $A = 0$ throughout the wall (Bloch wall solution). The solid lines are first-order phase boundaries, based on the numerically observed metastability (hysteretic behavior) of the two distinct solutions on either side of the boundary. The dashed line is a second-order phase boundary, again based on simulated annealing results. Also shown in Fig. 3 are phase boundaries for various values of the antiferromagnetic gradient coefficient $g_A$, indicating that the generic behavior is preserved away from the special case $g_A = 0$.

The AF cores in CMR materials, if they exist, would be insulating, as in the bulk. This would lead to high domain wall resistances, as is in fact observed. This resistance can be calculated in simple model of the band structure. Our model will be a classical spin $\vec{S}$ to represent the 3 localized $t_{2g}$ electrons on each Mn site, and a single tight-binding band to represent the current-carrying delocalized $e_{2g}$ electrons. The Hamiltonian for the $e_g$ electrons is then $H = -t\sum c_{is}^{\dagger} c_{js} - J\sum \vec{S}_i \bullet \vec{s}_i$. Here $i$ is a site label and $\vec{s}_i$ is the spin of the $e_g$ conduction electron.

$J$ is the Hund's rule coupling. The spins $\vec{S}_i$ of the $t_{2g}$ electrons are held fixed. We adopt a picture of the domain wall where $\vec{S}_i = M(x)\hat{z} + A(x)\hat{x}$, and the position dependence of $M$ and $A$ is given in Fig. 1. A Néel structure is assumed for the cubic lattice of Mn atoms.

We compute the spin-dependent (multichannel) transmission matrix numerically. From this, the resistance of the domain wall follows by using a generalized Landauer formula. The resistance as a function of wall width is shown in Fig. 4. The resistance is roughly exponential. This can be understood within a WKB tunneling picture. If the Fermi level $E_F$ of the metal lies in the gap of the insulator, then the wavefunction $\psi$ is damped in the wall: $\psi \sim \exp(x/x_o)$ in the wall, where $x_o = at/|E_G - E_F|$. Here $|E_G - E_F|$ is the smallest distance from the metallic Fermi level to a gap edge in the insulator.

The calculation presented here gives the conductance per unit area $g_w$ of a single domain wall. In experimental measurements [6], the excess electrical resistance of a sample with multiple magnetic domain walls is compared to that of the sample in the single-domain state. The ratio of these ($\Gamma$) is related through simple geometrical effects to the ferromagnetic domain width $d$ and the single-domain conductivity $\sigma_D$ as $\Gamma \approx \sigma_D / g_w d$. Thin films of Pr$_{0.67}$Sr$_{0.33}$MnO$_3$ experimentally show large domain wall resistivity [6], are strained by the substrate, with a Curie temperature lower than unstrained material. Strain has moved the sample closer to the antiferromagnetic phase boundary. In addition, the strain generates an out-of-plane magnetic anisotropy and corresponding dense domain structure. In this case, the ferromagnetic domain size is on the order of the film thickness, typically 200 Å in these thin films.

Experimentally the ratio $\Gamma$ is found to vary with temperature, but can be very large [6]. We can get an approximate upper bound for the wall conductance per unit area by noting that $g_w = (e^2/h) n^{2/3} \langle T \rangle$, where $n = 5.56 \times 10^{21}$/cm$^2$ is the electron density and $\langle T \rangle < 1$ is the appropriately averaged transmission coefficient. If we take a domain conductivity $\sigma_d = 10^4 /(\Omega-\text{cm})$ typical of manganite systems and $d = 200$ Å, we find $\Gamma > 0.4$. Thus the wall resistance is comparable to the domain resistance at small thicknesses, and Fig. 4 shows that the wall resistance can easily dominate the entire resistance of the film. This is due to the presence of a switchable metal-insulator transition in the boundary between two conductors.

The presence of local regions of insulator on the metallic side of the transition also suggests some unusual routes for the first-order transition not only in thin films but also in bulk samples. Let us focus first on the case of tetragonal symmetry, for then the topological character of the problem simplifies: there is only one type of wall, which separates only two possible equivalent domains. A standard scenario as the transition is approached from the ferromagnetic side would be nucleation of nearly spherical regions of insulator in a homogeneous metallic phase. However, the ferromagnet will always have domains. (Even in equilibrium, the dipole interaction enforces this.) As the system approaches the transition, the new phase could nucleate in the walls; the insulating regions would be topologically planar rather than spherical, and would traverse the entire sample. Two routes through the transition can now be envisioned. As their local free energy is lowered, (1) the walls could proliferate and eventually join; (2) the wall widths could increase continuously (as suggested by Fig. 1) resulting in coalescence. Scenario (2) would be the result if the only forces in the problem are those represented by Eq. (1) plus dipole forces. However, higher-order gradient and anisotropy terms can set a fixed length scale for the wall width [10], which would promote scenario (1).

If the symmetry is higher, say cubic, then several different types of walls are possible and complex networks may drive the transition [9].

This work was supported by the NSF Materials Theory program, Grant No. 0081039.

REFERENCES

[1] S. A. Kivelson and V. J. Emery, in *The Los Alamos Symposium 1993: Strongly Correlated Electronic Materials*, ed. K.S. Bedell, Z. Wang, D. Meltzer, A.V. Balatsky, and E. Abrahams (Addison-Wesley, NY, 1994), p. 619.

[2] See, e.g., J. M. Tranquada, J. Phys. Chem. Solids **59**, 2150 (1998).

[3] A. Himeta, T. Kato, and M. Ogata, Phys. Rev. Lett. **88**, 117001 (2002).

[4] C.H. Chen and S.-W. Cheong, Phys. Rev. Lett. **76**, 4042 (1996).

[5] M. Fath, S. Freisem, A.A. Menovsky, Y. Tomioka, J. Aarts, J.A. Mydosh, Science **285**, 1540 (1999).

[6] H.S. Wang and Qi Li, Appl. Phys. Lett. **73**, 2360 (1998); H.S. Wang, Qi Li, Kai Liu, and C.L. Chien, Appl. Phys. Lett., **74**, 2212 (1999); Qi Li and H.S. Wang, J. Superconductivity **14**, 231 (2001).

[7] J. F. Gregg, W. Allen, K. Ounadjela, M. Viret, M. Hehn, S. M. Thompson, and J. M. D. Coey, Phys. Rev. Lett. 77, 1580 (1996).

[8] P.M. Levy and S. Zhang, Phys. Rev. Lett. **79**, 51101580 (1997).

[9] J. Burgy, M. Mayr, V. Martin-Mayor, A. Moreo, and E. Dagotto, Phys. Rev. Lett. **87**, 277202 (2001); for a review of such theories, see E. Dagotto, T. Hotta, and A. Moreo, Physics Reports **344**, 1 (2001)

[10] L. Priyadko, S.A. Kivelson, V.J. Emery, Y. Bazaliy, and E.A. Demler, Phys. Rev. B **60**, 7541 (1999); O. Zachar, S.A. Kivelson, and V.J. Emery, Phys. Rev. B **57**, 1422 (1998)

[11] S. Satpathy, Z. Popovic, and F. Vukajlovic, Phys. Rev. Lett. **76**, 960 (1996); V. Anisimov, J. Zaanen, and O.K. Andersen, Phys. Rev B **44**, 943 (1991).

FIGURE CAPTIONS

Figure 1: Typical solutions for the spatial dependence of the ferromagnetic (*M*) and antiferromagnetic (*A*) order parameters at a wall between oppositely oriented ferromagnetic domains. The parameter *f* indicates the proximity of the material to the antiferro-ferro phase boundary in the bulk system.

Figure 2: Width of the antiferromagnetic domain-wall core as a function of proximity to the antiferro-ferro phase boundary for $\gamma = 2$. The inset indicates the logarithmic divergence for different values of the coexistence coefficient $\gamma$. The dashed portion of the curve indicates the regions where the ferromagnetic domain wall solution is of lower free energy.

Figure 3: Phase diagram of the system, showing regions in which the domain wall has an antiferromagnetic, insulating core. The three phase boundaries shown are for different values of the antiferromagnetic gradient coefficient.

Figure 4: Interface resistance of the domain wall as a function of wall width. The width is defined as the length over which the antiferromagnetic moment *A* is nonzero, measured in unit cells. The resistance grows roughly exponentially, with minor interference effects superposed. The calculations are for a tight-binding model with an electron density of $n = 5.56 \times 10^{21}/\text{cm}^2$, $J = 0.88$ eV, and $t = 0.27$ eV as estimated from band-structure calculations [11].

Figure 1:

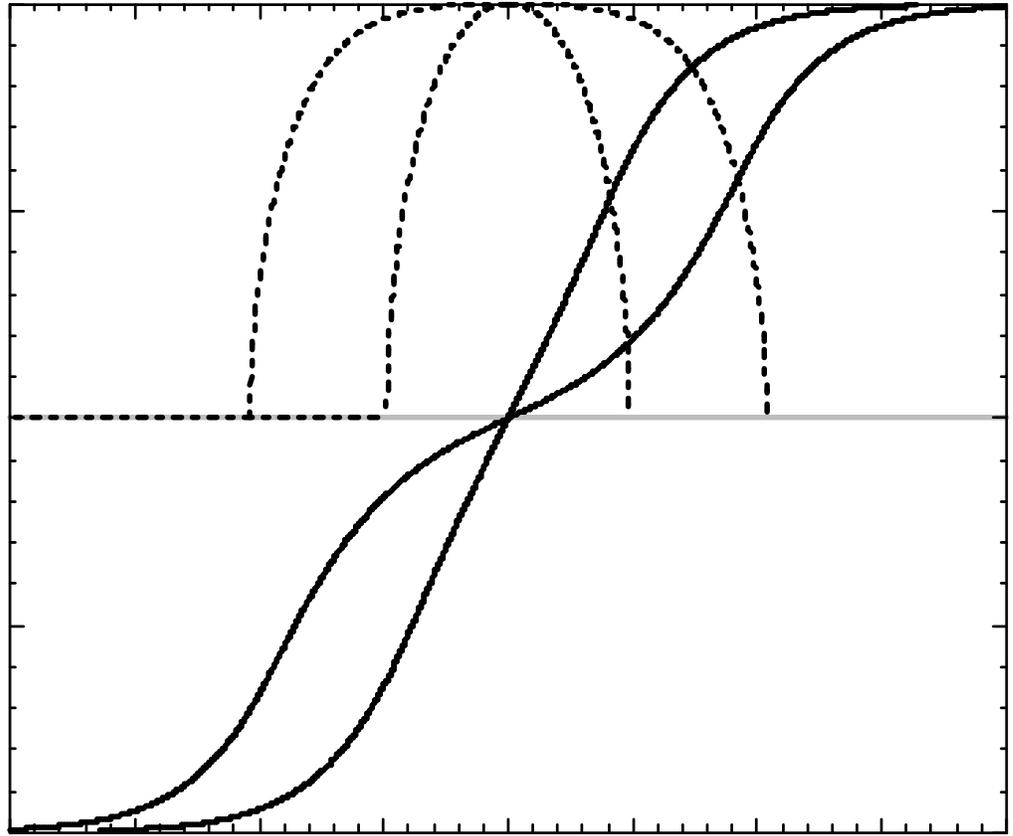

Figure 2:

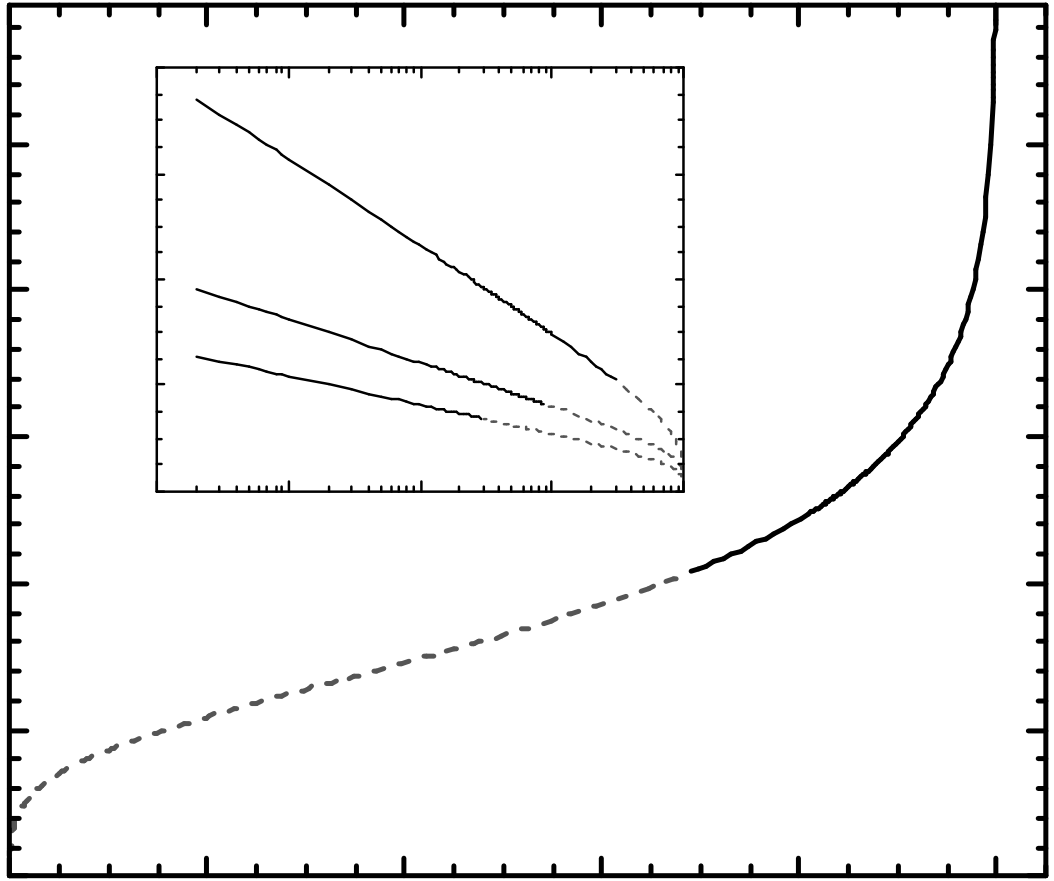

Figure 3:

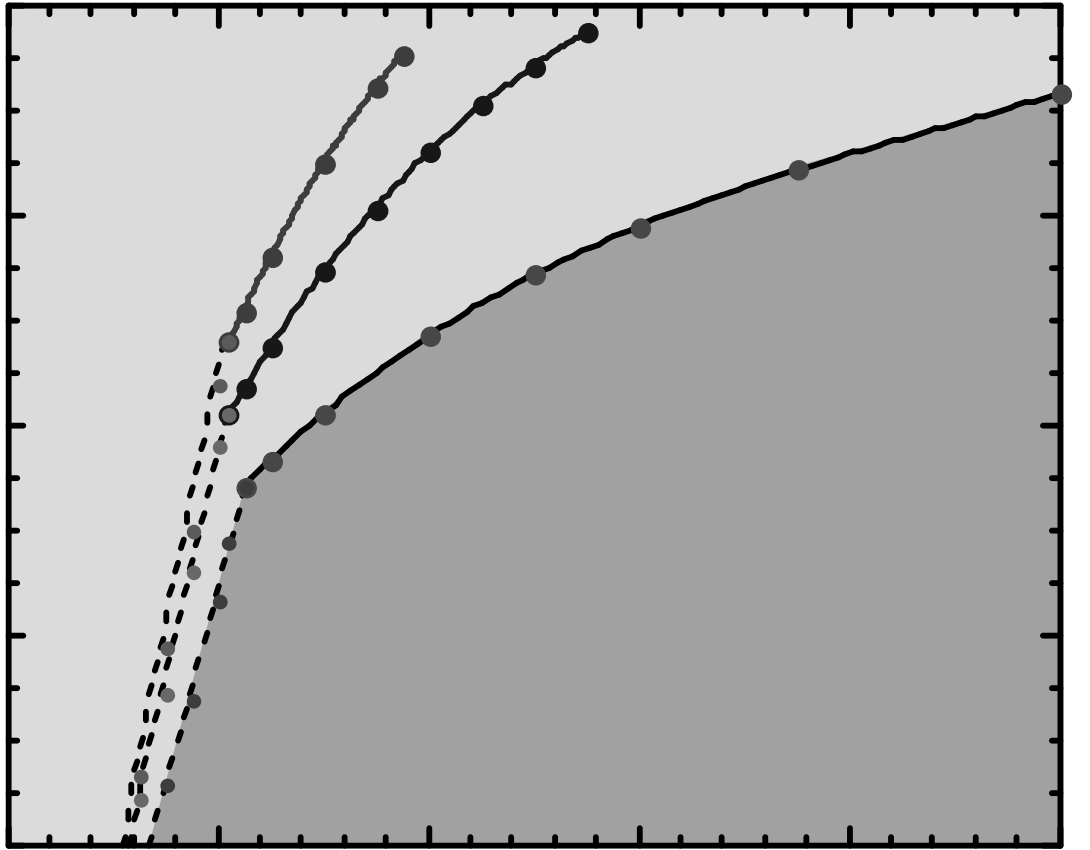

Figure 4:

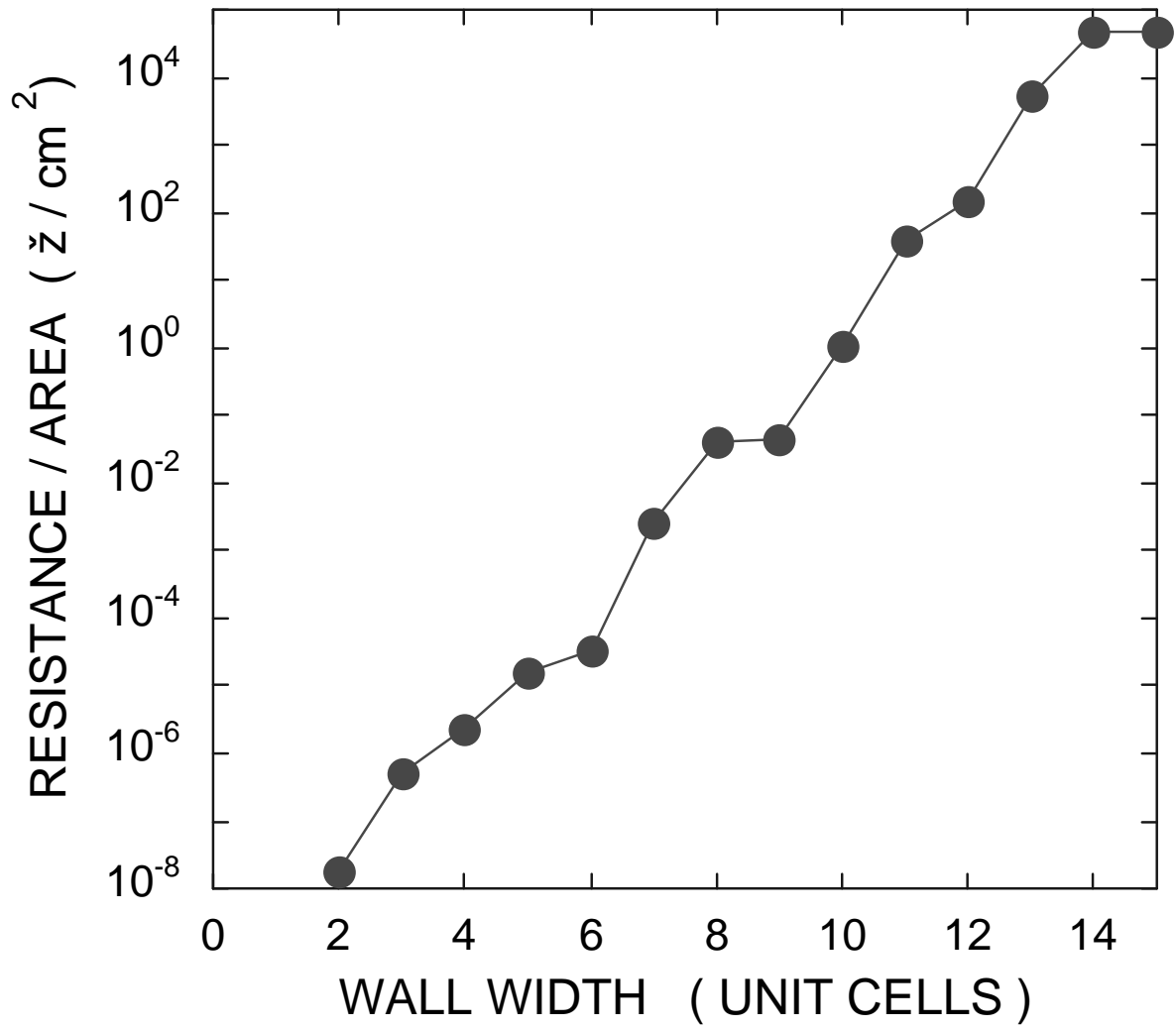